\documentclass[a4,11pt]{article}
\usepackage{amsmath, amssymb}
\usepackage[T1]{fontenc}
\usepackage{graphicx}
\usepackage{amsmath}
\usepackage{amssymb}%
\def \t {\tilde}
\def \si {\sigma}

\def \beg{\begin{eqnarray}}
\def \en{\end{eqnarray}}
\def \be*{\begin{eqnarray*}}
\def\e*{\end{eqnarray*}}
\def \di{\displaystyle}
\def\bit{\begin{itemize}}
\def \eit{\end{itemize}}

\def \w{\widehat}

\newtheorem{prop}{\sc Proposition}[section]
\newtheorem{lem}{\sc Lemma}[section]
\newtheorem{rem}{\sc Remark}[section]

\def \cqfd{\hspace*{14.6cm}$\blacksquare$}

\begin{document}

\title{{\itshape Testing for equality between two transformations of random variables }}

\author{Mohamed BOUTAHAR $^{\ast}$\thanks{$^\ast$Corresponding author,
IML. Luminy Faculty of Sciences.\, 163 Av.
de Luminy 13288 Marseille\ Cedex 9 \, e-mail: boutahar@univmed.fr.}  and Denys POMMERET $^{\dagger}$ \thanks{ $^\dagger $
IML. Luminy Faculty of Sciences.\, 163 Av. de Luminy 13288 Marseille  Cedex 9 e-mail: pommeret@univmed.fr.}}

\maketitle

\begin{abstract}

Consider two  random variables
contaminated by two unknown transformations.
 The aim of this paper  is to test  the equality of those transformations.
 Two cases are distinguished: first, the two  random variables have known distributions.
 Second,  they are unknown but observed before contaminations.
We propose a nonparametric test statistic based on empirical cumulative distribution functions.
Monte Carlo studies are performed to analyze the level and the power of the test.
An illustration is presented through a real data set.

\end{abstract}


\section{Introduction}

There exists an important literature concerning the deconvolution problem, when an unknown signal $Y$
is contaminated by a noise $Z$, leading to the
observed signal
\beg
\label{conv}
X&=&Y+Z.
 \en
 A major problem is to reconstruct the density of $Y$.
 Many authors studied the univariate problem when the noise $Z$
 has known distribution (see for instance Fan \cite{fan}, Carroll and Hall \cite{carhal}, Devroye \cite{dev}, or more recently Holzmann \emph{et al.} \cite{holz} for a review).
Bissantz \emph{et al.} \cite{biss}  proposed the construction of confidence bands for the density of $Y$ based on i.i.d. observations from (\ref{conv}).
The case where both $Y$ and $Z$ have unknown distributions is considered in Neumann \cite{neu},
Diggle and Hall \cite{dig}  or Johannes \emph{et al.} \cite{joh} among others.
When
the error density and the distribution of $Y$  have different characteristics the model can be identified as shown  in  Butucea and Matias \cite{but} and Meister \cite{mei}.
But
without information on  $Z$, the model suffers of identification conditions.
One solution is to assume  another independent sample is observed
from the measurement error $Z$ (as done in
  Efromovich  and  Koltchinskii \cite{efr} and   Cavalier and  Hengartner \cite{cav}).

 A more general  model than (\ref{conv}) occurs when the contaminated random variables are observed through a  transformation; that is, there exists  $g$ such that
\beg
\label{convgen}
X&=&g(Y+Z).
 \en
 When $g$ is known the problem is to estimate the distribution of $Y$, observing a sample from (\ref{convgen}).   An application of this model to fluorescence lifetime measurements is given in Comte and  Rebafka \cite{comt}. The authors developed an adaptative estimator that take into account the perturbation from the unknown  additive noise, and the
distortion due to the nonlinear transformation.

In this paper we consider a two sample problem of contamination that can be related to models (\ref{conv}) and (\ref{convgen}) as follows:  We assume that two contaminated random variables  are observed, say
$X$ and $\t{X}$,  which are transformations of two known, or observed, signals, that is:
\beg
\label{conv2}
X  =
g(Y), & &
\t{X}  =
\t{g}(\t{Y}),
\en
where $g$ and $\t{g}$  are continuous  monotone unknown functions.
Our purpose is to test
\beg
\label{hyptest}
 H_0: g =\t{g} & {\rm against} &
H_1: g\neq \t{g},
\en
based on  two i.i.d. samples
satisfying (\ref{conv2}).
The problem of testing (\ref{hyptest}) is of interest in many applications
when a signal is noised  in another way than the additive noise model (\ref{conv}).
We  will distinguish two important cases:
\bit
\item[\emph{Case 1}]
The distributions of $Y$ and $\t{Y}$ are known and we observe two samples reflecting  $X$ and $\t{X}$.
This situation may be encountered when two signals are controlled
in entry but observed with perturbations  in exit of a system.
\item[\emph{Case 2}]
The distributions of $Y$ and $\t{Y}$ are unknown and we first observe two independent samples based
on  $Y$ and $\t{Y}$, and then we observe
contaminated samples   $X$ and $\t{X}$ satisfying (\ref{conv2}).
This situation may be encountered when two unknown signals are observed both
in entry and  in exit of a system.
\eit
For both cases we construct a test statistics based on non parametric empirical estimators
of $g$ and $\t{g}$ and we adapt  a limit result on empirical processes due to Sen \cite{sen}.
Our test statistics are very easily implemented and
 we observe through simulations that they have a good power against various alternatives.
It is clear that  when  $H_0$   is not rejected;
  that is when the two noise functions are identical, it is then of interest to interpret the common estimation of ${g}$.
 We illustrate this point with a study of the Framingham dataset (see   Carroll \emph{et al.} \cite{car}, and more recently Wang and Wang \cite{wan}).

The paper is organized as follows: in Section 1 we consider the problem when the two
original signals have known distributions.  In Section 2 we relax the last assumption by assuming
unknown distributions but we observe  the two original signals
after and before perturbations. In Section 3 a simulation
study is presented and a real data set is analyzed.

\section{The test statistic}
\subsection{Case 1: the two signal distributions of $Y$ and $\t{Y}$ are known}
We consider $n$ (resp. $\t{n}$) i.i.d.  observations $X_1, \cdots, X_n$ (resp. $\t{X}_1, \cdots, \t{X}_{\t{n}}$)
from (\ref{conv2}). We assume that $Y$ and $\t{Y}$ are independent.
Write $F_Y$ and $F_{\t{Y}}$ the cumulative distribution functions of $Y$ and $\t{Y}$  respectively.
We assume that these functions are known and invertible.
We also write $F_X$ and $F_{\t{X}}$ the cumulative distribution functions of $X$ and $\t{X}$.
Also we assume that the  transformations $g$  and $\t{g}$  are monotone  and, without loss of generality, that they are increasing. Note that $g(y)= F^{-1}_X ( F_Y (y) )$ and  $\t{g}(y)= F^{-1}_{\t{X}} ( F_{\t{Y}} (y) )$. Hence a natural nonparametric  estimators of the contaminating
functions  are given by
\begin{eqnarray}
\widehat{g}(\cdot )=X_{( [ n F_{Y}(\cdot ) ] + 1   )} &\mbox{ and }&
\widehat{\tilde{g}}(\cdot)=\t{X}_{( [ \t{ n} F_{\t{Y}}(\cdot ) ] +1   )},
\label{hht}
\end{eqnarray}
where $X_{(i)}$  and   $\t{X}_{(i)}$ denote the $i$th order statistics, and $[x]$ denotes the integer part of the real $x$.
A fundamental theorem of Sen \cite{sen}  states the following convergence in distribution
\beg
\sqrt{n}\big( X_{( [n p]+1 ) } -F_{X}^{-1}(p)\big)\di \stackrel{D}{\rightarrow} {\cal N}\left( 0,\frac{p(1-p)}{f^{2}(F_{X}^{-1}(p))}\right) ,\
\forall p\in (0,1),
\label{co3}
\en
where $\stackrel{D}{\rightarrow}$ denotes the convergence in distribution, $f$ denotes the density of $X$ and ${\cal N}(m,\si^2)$ the Normal distribution with mean $m$ and variance $\si^2$.
We will need the following two standard assumptions:
  \bit
  \item
  $(A_1)$ there exists $a<\infty$ such that $n/(n+\t{n}) \rightarrow  a$
  \item
  $(A_2)$ $f>0$ and $f$ is ${\cal C}^k$, for some positive integer $k$.
  \eit

We deduce a first   result which is a main tool for the construction of the test statistic.
\begin{prop}
\label{prop1}\emph{ Let Assumption $(A_1)-(A_2)$ hold. Under $H_0$ we have
\beg
\sqrt{\di\frac{n\t{n}}{n+\t{n}}}\left(
\widehat{g}(y) -
\widehat{\t{g}}(y)
\right)
\di\stackrel{D}{\rightarrow}{\cal N}(0,\sigma^2(y)),\text{ as
} n \rightarrow \infty,  \t{n} \rightarrow \infty,
\label{nor}
\en
where
$$
\sigma^2(y) = (1-a)\frac{F_Y(y)(1-F_Y(y)) }{ f_X^{2}(g(y))} +a
\frac{F_{\t{Y}}(y)(1-F_{\t{Y}}(y))}{ f_{ \t{X}} ^{2}(\t{g}(y)) }
$$}
\end{prop}
{\sc Proof.} It follows directly from (\ref{co3}), replacing $p$ by
$F_Y(y)$ and  $F_{\t{Y} } (y)$ respectively.

\cqfd

We will estimate the variance $\si^2$ by using a nonparametric method. Consider a kernel
$K(\cdot)$, for instance the quartic kernel  defined by $K(y) = \frac{15}{16} (1-y^2)^2  {\bf 1} _{(-1,1)}(y)$, and  an associated bandwidth $h_n$. In the sequel,  we will set $K_{h_n}(y)=K(\frac{y}{h_n})$.
To avoid small values for  denominators in the estimation of the variance we use

\be*
\widehat{f}_X(y) & = & \max\left(\di \frac{1}{nh_n}\di\sum_{i=1}^{n}K_{h_n}(X_i-y), e_n\right)
\e*
 and
 \be*
\widehat{f}_{\t{X}} (y)   =   \max \left( \di \frac{1}{\t{n}h_{\t{n}}}\di\sum_{i=1}^{\t{n}}K_{h_{\t{n}}}(\t{X}_i-y),e_{\t{n}} \right),
\e*

where $e_n >0 $ and $e_n \rightarrow 0$ when $n$ tends to infinity. The estimator of $\si^2$ is then
\be*
\w{\si}^2(y)  & = & (1-a)\frac{F_Y(y)(1-F_Y(y)) }{ \w{f}_X^{2}(\w{g}(y))} +a
\frac{F_{\t{Y}}(y)(1-F_{\t{Y}}(y))}{ \w{f}_{ \t{X}} ^{2}(\w{ \t{g}}(y)) },
\e*
and we consider the statistic
\beg
T_1(y) & = &
\di\frac{n\t{n}}{n+\t{n}}\w{\si}(y)^{-2}\left(
\widehat{g}(y) -
\widehat{\t{g}}(y)
\right)^2.
\label{stat1}
  \en

  \begin{prop}
 \label{prop3}\emph{
 Let Assumptions $(A_1)$-$(A_2)$ hold. If $h_n\simeq n^{-c_{1}}$, $e_{n}\simeq n^{-c_{2}}$ for some positive
constants $c_{1}$ and $c_{2}$ such that $\frac{c_{2}}{k}<c_{1}<\frac{1}{1+2k}$,
  then under $H_0$, when $n\to \infty, \t{n}\to\infty$,   we have for all $y$:
 \be*
 T_1(y)  \stackrel{D}{\rightarrow}  Z  ,
 &  &
 \e*
 where $Z$ is Chi-squared distributed with one degree of freedom}.
 \end{prop}

{\sc Proof.}
We need the fundamental Lemma (see  H\"{a}rdle \cite{hard}):
\begin{lem}
\label{lem1}
$$
 \sup_{y\in\mathbb{R}}|\hat{f}^2(y)-f^2(y)|=\mathcal{O}_{p}
\left(h_n^{2k}+\frac{\log{n}}{{{n}}h_n}\,\right).$$
\end{lem}
We can write
\be*
\w{\si}^2(y) & = & \di\frac{u(y)}{\w{f}^2_X(\w{g}(y))}
+ \di\frac{v(y)}{\w{f}^2_{\t{X}}(\w{\t{g}}(y))},
\e*
where
$u(y) = (1-a)F_Y(y)(1-F_Y(y))$ and $v(y) = a F_{\t{Y}}(y)(1-F_{\t{Y}}(y))$. Using  Taylor expansion there exist $A$ and $B$ such that
\be*
\w{\si}^2(y) & = & \si^2(y) +
\bigg(\w{f}^2_X(\w{g}(y))-f^2(g(y))\bigg)\bigg(\di\frac{-1}{A^2}\bigg)
+
\bigg(\w{\t{f}}^2_X(\w{\t{g}}(y))-\t{f}^2(\t{g}(y))\bigg)\bigg(\di\frac{-1}{B^2}\bigg),
\e*
 with
 \be*
 \di\frac{1}{A^2} \leq \di\frac{1}{e_n^2}
 &&
 \mbox { \,\, and \,\, } \di\frac{1}{B^2} \leq \di\frac{1}{e_{\t{n}}^2}.
 \e*
 Then, from Lemma \ref{lem1} we get
\be*
 \w{\si}^2(y) & = &
 \si^2(y) +
 o_{P}(1),
\e*
by assumption
and the result follows from Proposition \ref{prop1}.

\cqfd

\subsection{Case 2: the two signal distributions $Y$ and ${\t{Y}}$ are unknown}
We consider $n_x$ (resp. $\t{n_x}$) i.i.d.  observations $X_1, \cdots, X_{n_x}$ (resp. $\t{X}_1, \cdots, \t{X}_{\t{n_x}}$)
and $n_y$ (resp. $\t{n_y}$) i.i.d. observations $Y_1, \cdots, Y_{n_y}$ (resp. $\t{Y}_1, \cdots, \t{Y}_{\t{n_y}}$) from (\ref{conv2}).
Put
\be*
N=n_x n_y/(n_x+n_y) & {\rm and }  &  \t{N}=\t{n_x}\t{n_y}/(\t{n_x}+\t{n_y}).
\e*
The two samples $Y_1, \cdots, Y_{n_y}$ and  $\t{Y}_1, \cdots, \t{Y}_{\t{n_y}}$ can be viewed as two independent training sets which permit to estimate the initial densities of the signals before perturbations.
Again we want test $H_0: g = \t{g}$.
We now estimate  $g$ and $\t{g}$ by

\begin{eqnarray}
\widehat{g}(\cdot )=X_{( [ n_x \w{F}_{Y}(\cdot )] )} & \mbox{ and }&
\widehat{\tilde{g}}(\cdot)=\t{X}_{( [ \t{ n}_x \w{F}_{\t{Y}}(\cdot )]  )},
\label{estimh}
\end{eqnarray}
 where
\begin{eqnarray}
\w{F}_{Y}(y)=\frac{1}{n_y}\sum_{i=1}^{n_y} {\bf 1}_{ \{Y_{i}\leq y\}} &\mbox{ and }&
\w{F}_{\t{Y} }(y)=\frac{1}{\t{n}_y}\sum_{i=1}^{\t{n}_y}{\bf 1}_{ \{\t{Y}_{i}\leq y\}},
\label{emp}
\end{eqnarray}
are the empirical distribution functions of $Y$   and   $\t{Y}$ respectively.
We assume
that
\be*
\lim n_x/(n_x+{n_y}) = a < \infty, && \lim \t{n_x}/(\t{n_x}+\t{n_y})= \t{a} < \infty,
\e*
and we make the following   assumption, extending Assumption (A1):
  \bit
  \item
  $(A_3)$ there exists $b<\infty$ such that $N/(N+\t{N}) \rightarrow  b$. 
  \eit
We can  extend Proposition \ref{prop1}  as follows:
\begin{prop}
\label{prop4}\emph{
Let Assumption $(A_2)-(A_3)$ hold. Under $H_0$ we have
\beg
\sqrt{\di\frac{N\t{N}}{N+\t{N}}}\left(
\widehat{g}(y) -
\widehat{\t{g}}(y)
\right)
\di\stackrel{D}{\rightarrow}{\cal N}(0,\sigma^2(y)),\text{ as
} N \rightarrow \infty,  \t{N} \rightarrow \infty,
\label{nor}
\en
where
\beg
\sigma^2(y) = (1-b)\frac{F_Y(y)(1-F_Y(y)) }{ f_X^{2}(g(y))} + b
\frac{F_{\t{Y}}(y)(1-F_{\t{Y}}(y))}{ f_{ \t{X}} ^{2}(\t{g}(y)) }.
\label{sig}
\en}
\end{prop}

{\sc Proof.}
We first show that
\be*
U=\sqrt{ \di\frac{n_x n_y}{n_x+n_y}}\left( \widehat{g}(y) - g(y) \right)
\di\stackrel{D}{\rightarrow}{\cal N}(0,\sigma_1^2(y)),\text{ as
} n_x \rightarrow \infty,  {n_y} \rightarrow \infty,
\e*
where  $$
\sigma_1^2(y) = \frac{F_Y(y)(1-F_Y(y)) }{ f_X^{2}(g(y)) }.$$

For that write
\be*
\w{g}(y) -g(y) &=& \w{G}(y)  + G(y),
\e*
where $\w{G}(y)=\w{g}(y)-F_X^{-1}(\w{F}_Y(y))= X_{( [ n_x \w{F}_{Y}(x )] )} - F_X^{-1}(\w{F}_Y(y))$  and
$G(y) = F_X^{-1}(\w{F}_Y(y))-g(y)$.
By the delta method we get

\be*
n_y^{1/2}G(y) & \to & {\cal N}(0,\si_1^2(y)).
\e*

Then we decompose the characteristic function
\be*
E\left(e^{iu U}\right) & = &
E \left(e^{iu n_{x,y} G} E (e^{iun_{x,y} \widehat{G}  }| {\bf Y})\right),
 \e*

  where $n_{x,y}= \sqrt{\di\frac{n_xn_y}{n_x+n_y}}$ and  ${\bf Y}$ stands for the vector of observation $Y_1, \cdots, Y_{n_y}$.

Since these functions are bounded we get:
\be*
\lim_{n_x\to \infty, n_y\to \infty}  E(\exp(iuU)) & = &
E\left(\lim_{n_x\to \infty, n_y\to \infty} e^{iu n_{x,y} G} \lim_{n_x\to \infty, n_y\to \infty} E\left(e^{iu n_{x,y} \w{G} }|{\bf Y}\right)\right)
\\
& = &
 E\left(e^{iu\sqrt{a} Z} \lim_{n_y\to \infty}e^{-\frac{1}{2}(1-a) \w{\si}_1^2(y)}\right),
\e*
where $Z \sim {\cal N}(0,\si_1^2(y))$ and $\w{\sigma}_1^2(y) = \frac{\w{F}_Y(y)(1-\w{F}_Y(y))}{f_X^{2}(g(y))}$.
We finally obtain
\be*
\lim_{n_x\to \infty, n_y\to \infty}  E(\exp(i u U)) & = &
\exp(-1/2 u^2 \si^2_1(y)).
\e*

Similarly, writing
\be*
\t{U}& = & \sqrt{\di\frac{\t{n}_x\t{n}_y}{\t{n}_x+\t{n}_y}}\left( 
\widehat{\t{g}}(y) -
\t{g}(y)
\right),
\e*
we obtain that
\be*
\t{U}
&\di\stackrel{D}{\rightarrow}&
{\cal N}(0,\t{\sigma}_1^2(y)),\text{ as
} \t{n}_x \rightarrow \infty,  \t{n}_y \rightarrow \infty,
\e*
with $$
\t{\sigma}_1^2(y) = \frac{F_{\t{Y}}(y)(1-F_{\t{Y}}(y))}{f_{\t{X}}^{2}(\t{g} (y)) }.$$

Finally, combining these two convergences with the equality
$\t{g}=g$ under $H_0$ we complete the proof.

\cqfd

As previously we can estimate $\sigma^2(y)$ in (\ref{sig}) by a nonparametric estimator
\be*
\w{\si}^2(y)  & = & (1-b)\frac{\w{F}_Y(y)(1-\w{F}_Y(y)) }{ \w{f}_X^{2}(\w{g}(y))} +b
\frac{\w{F}_{\t{Y}}(y)(1-\w{F}_{\t{Y}}(y))}{ \w{f}_{ \t{X}} ^{2}(\w{ \t{g}}(y)) },
\e*
where
$\w{F}_Y$ and $\w{F}_{\t{Y}}$  are the empirical distribution functions of $Y$   and   $\t{Y}$ given by (\ref{emp}).
Our test statistic is given by
\beg
T_2(y) & = &
\di\frac{N\t{N}}{N+\t{N}}{\w{\si}}^{-2}(y)\left(
\widehat{g}(y) -
\widehat{\t{g}}(y)
\right)^2.
\label{stat2}
  \en
  We can now generalize Proposition \ref{prop3} as follows.
  \begin{prop}
 \label{prop5}\emph{
Let Assumptions $(A_2)$-$(A_3)$ hold. If $h_{n}\simeq n^{-c_{1}}$, $e_{n}\simeq n^{-c_{2}}$ for some positive
constants $c_{1}$ and $c_{2}$ such that $\frac{c_{2}}{k}<c_{1}<\frac{1}{1+2k}$,
  then under $H_0$,
 when $ N\to \infty, \t{N}\to\infty$,   we have:
 \be*
 T_2 \stackrel{D}{\rightarrow}  Z  ,
 &   &
 \e*
 where $Z$ is Chi-squared distributed with one degree of freedom.}
  \end{prop}

{\sc Proof.}
We combine the proof of Proposition \ref{prop1} with the fact that
$\w{F}(1-\w{F})$ is bounded to get
\be*
 \w{\si}^2(y) & = &
 \si^2(y) +
 o_{P}(1),
\e*
and we conclude by Proposition \ref{prop4}.

\cqfd

\subsection{Behaviour of the tests under $H_1$}

We study convergence properties of the tests $T_1$ and $T_2$ under some alternatives
 \begin{prop}
 \label{prop6}
 \,\, \hspace{1cm } \,\, \\
 
 a. General alternatives. \,  \\
 Consider the test statistics  $T_1$ and $T_2$, then for all $y$ such that
 $g(y) \not = \t{g}(y),$ we have
 \begin{eqnarray*}
 T_1(y)& \stackrel{P}{\rightarrow} + \infty \mbox { and }  T_2(y) \stackrel{P}{\rightarrow} + \infty,
 \end{eqnarray*}
  where $\stackrel{P}{\rightarrow}$ denotes the convergence in probability. \\
  
 b. Local alternatives.  \\
Let us denote  $m=\frac{n \t{n}}{n+\t{n}}$ or  $m=\frac{N \t{N}}{N+\t{N}}$ according to whether if the test statistic  $T_1$ or $T_2$ is used and consider the local alternatives
  $$H_{l1}: \t{g}(y)= g(y)+ \frac{k(y)}{m^\beta},$$
  then under $H_{l1}$  and when $n\to \infty$, $\t{n}\to\infty$, $N\to \infty$, $\t{N}\to\infty$ we have for all $y$: \\
  i. If $\beta > 1/2$  then 
 \be*
 T_1(y)   \stackrel{D}{\rightarrow}  Z    \mbox{ and }
      T_2(y)   \stackrel{D}{\rightarrow}  Z
 &  &
 \e*
 ii.  If $\beta =1/2$ then 
  \be*
 T_1(y)   \stackrel{D}{\rightarrow}  Z_k   \mbox{ and }      T_2(y)   \stackrel{D}{\rightarrow}  Z_k
 &  &
 \e*
  iii. If $\beta <1/2$ then 
  \be*
 T_1(y) \stackrel{P}{\rightarrow} + \infty \mbox { and }  T_2(y) \stackrel{P}{\rightarrow} + \infty
       &  &
 \e*
  where $Z$ is Chi-squared distributed with one degree of freedom
 and $Z_k$ is a decentred Chi-squared distributed with one degree of freedom and parameter $k(y).$ 
  \end{prop}
The proof of this proposition is straightforward and hence is omitted. \cqfd

\begin{rem}
Estimators  $\w{g}$ (resp.  $\widehat{\tilde{g}}$ ) are computed  from
$(X_1, \cdots, X_{n_x})$ and $(Y_1, \cdots, Y_{n_y})$  (resp. $(\t{X}_1, \cdots, \t{X}_{\t{n_x}})$ and  $(\t{Y}_1, \cdots, \t{Y}_{\t{n_y}})$ ).
Under the null $H_0$ there are two different ways to construct a common estimator of $g$. First we can consider the aggregate estimator
\begin{eqnarray}
 \w{g}_{0}& =&   \frac{ (n_x+n_y) \w{g} + (\t{n_x}+\t{n_y})\widehat{\tilde{g}} }{n_x+n_y+\t{n_x}+\t{n_y}},
 \label{agreg}
 \end{eqnarray}
and, second, another   estimator can be construct  by aggregating the samples.

\end{rem}

\section{Simulations and data study}

For all empirical powers or empirical levels we carry out  experiments of $10000$ samples and we use three different sample sizes: $n=50$, $ n=100$,  and $n=500$.
For each replication we compute the statistics $T_{1}(y)$ and $T_{2}(y)$  given by (\ref{stat1}) and (\ref{stat2}), where $y$ is chosen randomly following a standard normal distribution.
\subsection{Study of the empirical levels}

We will denote by ${\cal N}(0,1)$ the standard normal distribution with mean zero and variance 1.
  We first consider the case where $Y_t$  and $\t{Y}_t$  are independent and  ${\cal N}(0,1)$ distributed. The bandwidth is chosen as $h_n=n^{-1/2}$ and the trimming as
  $e_n = n^{-1/5}$.

\paragraph{Empirical level}
 To study the empirical levels   of $T_1$ and $T_2$ we choose
 $$ g(y) = \t{g}(y)=  \exp\{(y+3)/(y+5)\},$$
and we fix a theoretical level  $\alpha=5\%$. Table \ref{tab1}  shows empirical levels of the test under $H_0$. It can be seen  that both statistics $T_{1}$ and $T_2$ provide levels close to the asymptotic value.

\begin{table}[H]
\caption{Empirical levels of $T_1$  and $T_2$ (in \%) for a theoretical level $\alpha = 5\%$ .}
\label{tab1}
\begin{center}
\begin{tabular}{|c|c|c|c|}
\hline
& $n=50$ & $n=100$ & $n=500$    \\
\hline
$T_1$   & 3.9  & 4.75 & 5.49   \\
\hline
$T_2$  &  4.68 &    5.52  &  5.42  \\
\hline
\end{tabular}
\end{center}
\end{table}

 \subsection{Study of the empirical powers}

 We consider the model where  $Y_t$  and $\t{Y}_t$  are  independent and ${\cal N}(0,1)$ distributed.
 To study the empirical powers of $T_1$ and $T_2$ we consider  $g(y) =   \exp((y+3)/(y+5))$
 and the four following transformations:
 \be*
&& \t{g}_1(y)= \exp((y+3)/(y+5)) + 1 , \t{g}_2(y)= 2\exp((y+3)/(y+5)), \\  &&\t{g}_3(y)= -(y+11)/(y+5), \t{g}_4(y)= 4y + 5,
 \e*
 and we also study  local alternatives by considering:
$$ \t{g}_5(y)= g(y) + \frac{ 2(y+5) } {n^\beta}.$$
Tables \ref{tab3}-\ref{tab4} present empirical powers for $T_1$ and $T_2$ under fixed and local alternatives, respectively, for a theoretical level $\alpha$ equal to $5\%$.
From Table \ref{tab3} it appears that the knowledge of the probability densities of $Y$ and $\t{Y}$ allows to  have more stable statistics that detect more easily the departure from  the null hypothesis. Then  the test statistic $T_1$ provides better power, particularly for smallest sample size. The test statistic $T_2$ has a low empirical power
for $n=50$; but when the sample size $n$ increases, the empirical power of $T_2$ is similar  to that of $T_1$.
Table \ref{tab4} indicates that  $T_1$ and $T_2$ provide good power  for  $\beta\leq 1/2$.
 For $\beta > 1/2$  the power  converges to the theoretical level $\alpha$; this is in accordance with the theoretical result stated in Proposition \ref{prop6}.

\begin{table}[H]
\caption{Empirical powers of $T_1$  and $T_2$  (in \%) for a theoretical level $\alpha = 5\%.$}
\label{tab3}
\begin{center}
\begin{tabular}{|l|l|l|l|l|}
\hline
   & $T_1$ & $T_2$ & $T_1$ & $T_2$ \\
\hline
 & $\t{g}_1  $ & $\t{g}_1  $  & $\t{g}_2   $ & $\t{g}_2   $  \\
\hline
$n=50$    &  99.98    &  99.58       &        99.81  &    98.17   \\
\hline
$n=100$   &  99.99    &  99.66       &        99.91   &    98.17         \\
\hline
$n=500$    &    100    &   99.69      &         99.96   &    98.47
       \\
\hline
\hline
   & $T_1$ & $T_2$ & $T_1$ & $T_2$ \\
\hline
 & $\t{g}_3   $ & $\t{g}_3   $  & $\t{g}_4  $ & $\t{g}_4  $  \\
\hline
$n=50$      &  100    &   100    &   78.59      &    71.47     \\
\hline
$n=100$    &   100   &    100   &    84.31     &     78.41       \\
\hline
$n=500$    &     100  &    100   &    92.42    &    92.07
       \\
       \hline
\end{tabular}
\end{center}
\end{table}

\begin{table}[H]
\caption{ Empirical powers of $T_1$  and $T_2$  (in \%) for a theoretical level $\alpha = 5\%$ under local alternative $\t{g}_5$.}
\label{tab4}
\begin{center}
\begin{tabular}{|l|l|l|l|l|l|l|}
\hline
& $T_{1}$ & $T_{2}$ & $T_{1}$ & $T_{2}$ & $T_1$ & $T_2$  \\
\hline
& $ \beta= 1/4 $ & $   \beta= 1/4   $  &  $ \beta= 1/2 $ & $   \beta= 1/2 $   &  $ \beta= 4 $ & $   \beta= 4 $     \\
\hline
$n=50$     &     99.85         &   97.06      &   99.64    &    96.90   &     4.19   &   4.77        \\
\hline
$n=100$    &     99.85        &   97.30     &      99.71   &    97.02   &     4.77      &  5.72          \\
\hline
$n=500$    &     99.94       &    97.85      &     99.82     &   97.29   &     5.36    &   5.28        \\
\hline
\end{tabular}
\end{center}
\end{table}

 \subsection{Real example: Framingham  data}

We consider the Framingham Study on coronary heart disease described by Carroll et al. \cite{car}. The data consist of measurements of systolic blood pressure (SBP) obtained at two different examinations in 1,615 males on an 8-year follow-up. At each examination, the SBP was measured twice for each individual.
The four variables of interest are: \\
$Y=$  the first SBP at examination 1, \\
$\t{Y}=$  the second SBP at examination 1, \\
$X=$  the first SBP at examination 2, \\
$\t{X}=$  the second SBP at examination 2.

Our purpose is   to examine whether the distribution of the SBP changed during time, and which type of transformation it underwent. Following our notations, we will study the transformation between the distributions of $Y$ and $X$ and also the one between the distributions of $\t{Y}$ and $\t{X}.$ Then we assume that $X=g(Y)$ and $\t{X}=\t{g}(\t{Y}).$

\begin{table}[H]
\caption{ Descriptive statistics of Framingham data}
\label{tab11}
\begin{center}
\begin{tabular}{|l|l|}
\hline
   \, & \,   \\
 $Y$ & $X$  \\
 \hline
 Min. 1st Qu.  Median    Mean 3rd Qu.    Max. &   Min. 1st Qu.  Median    Mean 3rd Qu.  Max. \\
 80.0   \,120.0  \,\,\,  130.0   \,\,\,\, 132.8  \,\, 142.0   \,\, 230.0  &
   88.0 \,  118.0  \,\,\, 128.0  \,\,\, 131.2  \,\, 142.0  \,\, 260.0 \\ \\
 Var.   \,\,\,\, Skewness.  Kurtosis.  KS. &  Var.   \,\,\,\,  Skewness.  Kurtosis. \,\, KS. \\
419.12  \,1.27 \,\,\,\, \,\,\,\, \,\,\,\,7.79 \,\,\,\,\,\,\,\,\,\,\,\, 0.0119 &   439.11 \,  1.39 \,\,\,\, \,\,\,\, \,\,\,\,   6.65 \,\,\,\,\,\,\,\,\,\,\,\,  0.1125  \\
\hline
  \, &  \,  \\
 $\t{Y}$ & $\t{X}$  \\
 \hline
 Min. 1st Qu.  Median    Mean 3rd Qu.    Max.  &   Min. 1st Qu.  Median    Mean 3rd Qu.    Max. \\
 75.0  \, 118.0 \,\,\,  128.0 \,\,\,  130.2 \,\,  140.0 \,\,  270.0    & 85.0 \,  115.0\,\,\,   125.0  \,\,\, 128.8   \,\,138.0  \,\,  270.0   \\\\
 Var.   \,\, \,\,\,\, Skewness.  Kurtosis. KS. &  Var.   \,\,\,\, Skewness.  Kurtosis.  KS. \\
409.97 \,\,\, 1.46 \,\,\,\, \,\,\,\, \,\,\,  7.25  \,\,\,\,\,\,\,\,\,\,\,\,0.1171  &   410.21\,    1.47 \,\,\,\, \,\,\,\, \,\,\,\,  7.10\,\,\,\,\,\,\,\,\,\,\,\,   0.1117 \\
\hline
\end{tabular}
\end{center}
\end{table}

\begin{figure}
\includegraphics[width=14cm, height=10cm, keepaspectratio=true]{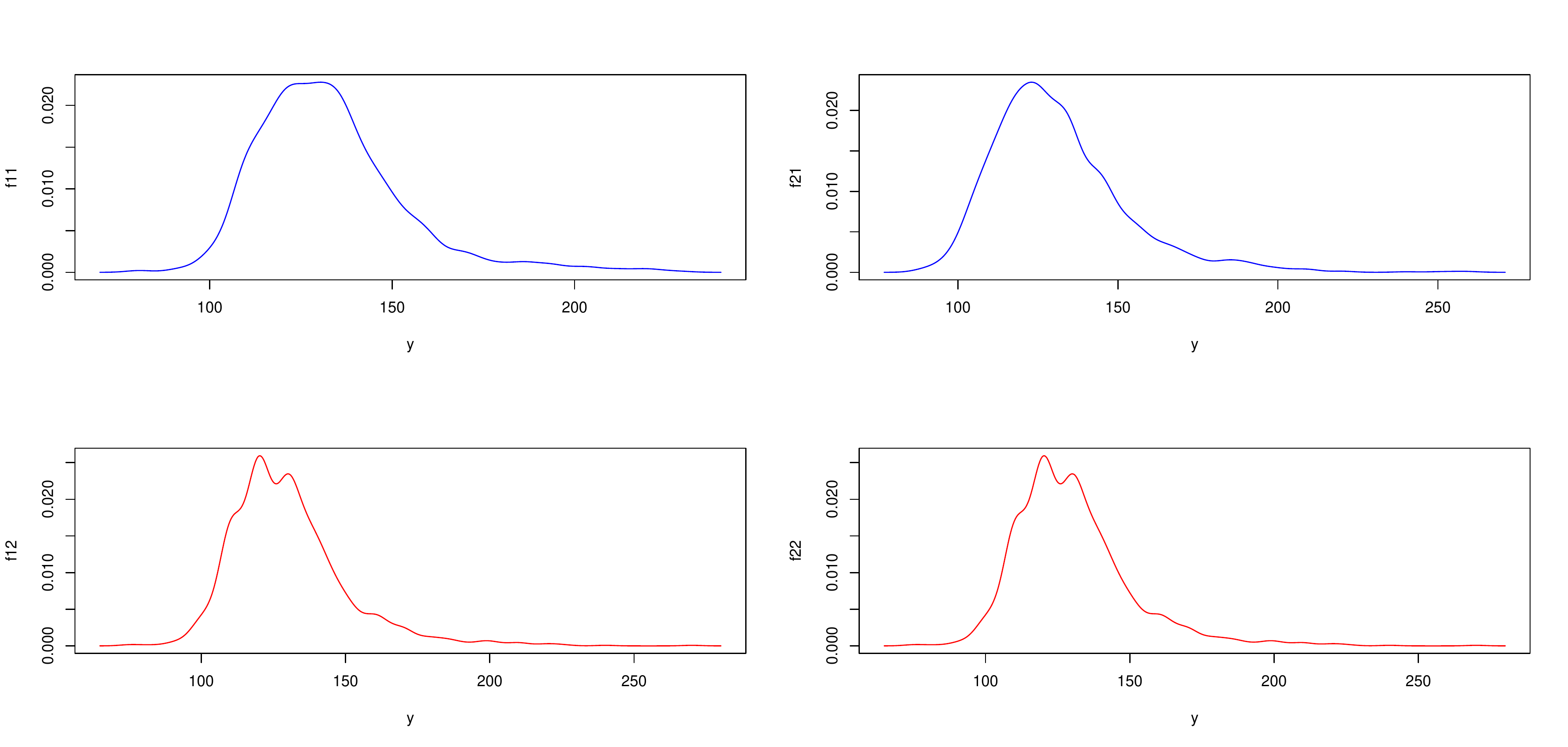}
 \caption{Kernel estimates of the probability densities of  $X,Y,\t{X},\t{Y}$.
 In the top panel : f11  (resp. f21) is the Kernel estimate of the density of $Y$ (resp. of $X$). In the bottom panel :  f12  (resp. f22) is the Kernel estimate of the density of $\t{Y}$ (resp. of $\t{X}$).}
\label{fig1}
 \end{figure}

Table \ref{tab11} indicates  that all the distributions of $X$, $Y$, $\t{X}$ and $\t{Y}$ are skewed to the right and are leptokurtic, $KS$ is the Kolomogorov-Smirnov statistic, the associated p-values are lesser than $2.210^{-6}$ and hence
the normality assumption is strongly rejected.
Figure \ref{fig1} represents nonparametric estimations of the probability densities of  $X,Y,\t{X}$, and $\t{Y}$.

 \begin{figure}
\includegraphics[width=14cm, height=10cm, keepaspectratio=true]{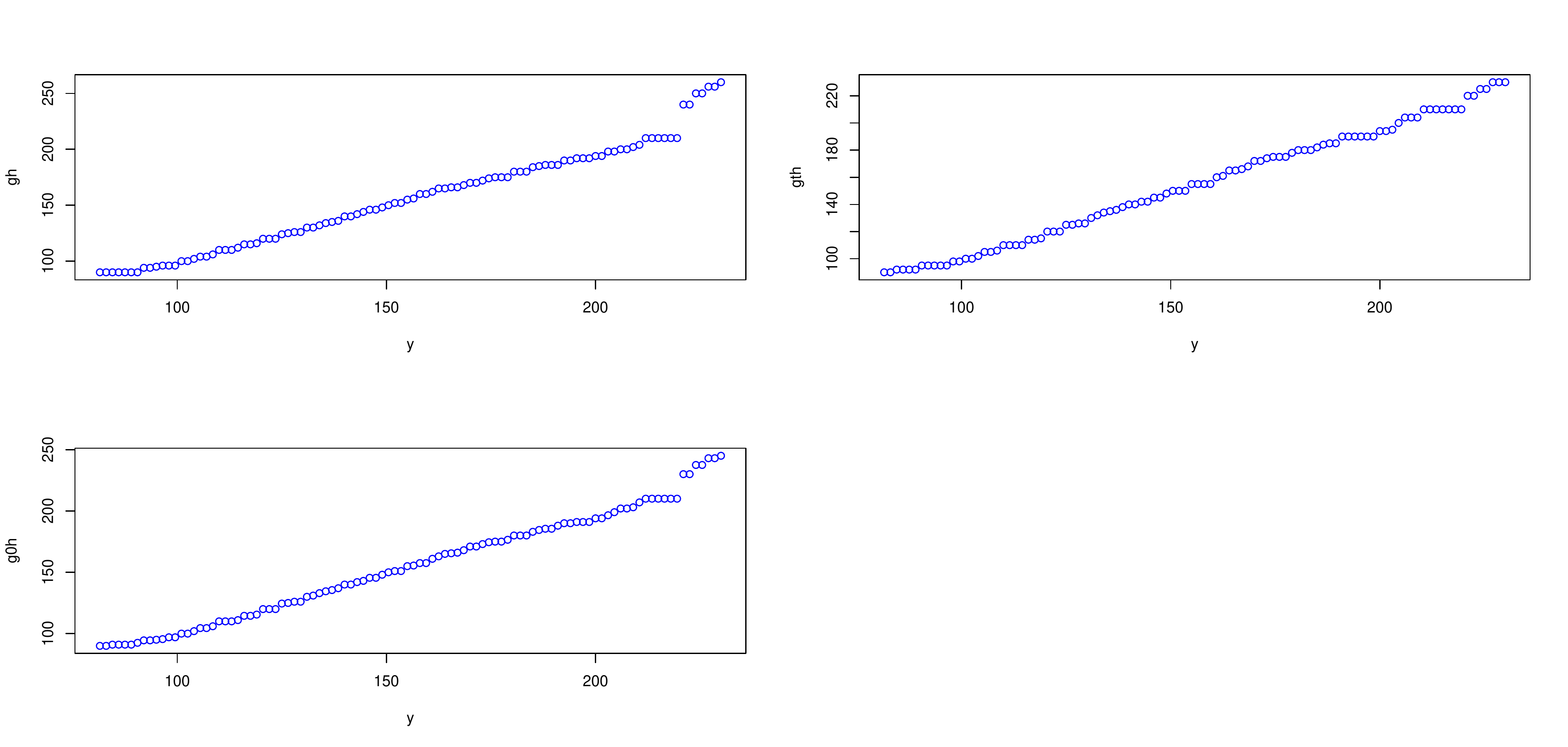}
\caption{Nonparametric estimators of $g$ and $\t{g}$ and the aggregated estimator on the interval $[c,d]$: $gh$ (resp. $gth$ and $g0h$) denotes $\w{g}$ (resp. $\w{\t{g}}$ and $\w{g_0}$).   }
\label{fig2}
 \end{figure}

From Figure \ref{fig1} it seems that the distributions  of the  variables $Y$  and $X$  have a similar shape. However, from Table \ref{tab11} we observe  a noticeable decrease in the mean and an increase in the variance. Based on the nonparametric estimators given in Figure \ref{fig2}   we can postulate that only the location and the scale are affected by time, therefore, the transformation  $g$ is linear; that is,
$ g(y)= ay +b.$
Similarly the distributions of the  variables $\t{Y}$  and $\t{X}$  can be linked by  $ \t{g}(y)= \t{a} y + \t{b} .$
The functions $g$, $\t{g}$ 
are estimated on the interval $[c,d]$
where $c= \max( \min(Y_i), \min(\t{Y}_j))$ and $d= \min( \max(Y_i), \max(\t{Y}_j))$. These functions are estimated on the grid $y_i = c + (d-c)i/M$,  for a given $M$.

By applying our test we obtain a p-value very close  to 1, and hence we can consider that $g=\t{g}$.

In Figure \ref{fig2}  we observe that all the estimators $\w{g}$, $\w{\t{g}}$ and $\w{g}_0$ are approximately linear on the interval $[c,d]$, however in the border (near $c$ and $d$) the approximation is not good. One can observe that they are constants  on regions where there are not enough observations. Therefore, to compute the linear approximation of these estimators we consider only the $y_i$ belonging to the interval $[100,200]$.

The ordinary least squares based on $(y_i, \w{g}(y_i) )$, $(y_i, \w{\t{g}}(y_i) )$ and  $(y_i, \w{g_0}(y_i) )$, $y_i \in [100,200]$, $1\leq i \leq 50 $ yields
$$ \w{g}(y)= 0.9877 y +  0.7035,  \w{\t{g}}(y)= 0.9857  y + 0.8335  \mbox { and }  \w{g_0}(y)= 0.986y +0.7685$$
By using a parametric approach, i.e. $\w{g}_p(y)=ay+b$, where  $a= \mbox{cov}(X,Y)/ \mbox{var}(Y),  b = \overline{X} - a \overline{Y},$
we obtain the following estimators

$$ \w{g}_p(y)= 0.760 y + 33.075,  \w{\t{g}}_p(y)= 0.726 y +36.730,$$
and the common aggregate parametric estimator is given by
$$ \w{h_{p,0}}(y)= 0.744 y + 34.787.$$
To compare the parametric and the nonparametric approaches, we consider the aggregate estimators and we compare the predicted values for the two first moments of $X$ and $\t{X} $ with those observed. The predictions of $X$ ( resp. of $\t{X}$) are computed by using the observed moments of $Y$ (resp. of $\t{Y}$) and the common transformation.  Using the parametric approach we get
 \begin{eqnarray*}
 \w{m}_X= 0.744 m_Y + 34.787
 = 133.590
 \\
  \w{Var}_X = (0.744)^2 Var(Y)
  = 232.145.
  \end{eqnarray*}
The nonparametric approach  yields
 \begin{eqnarray*}
 \w{m}_X = 0.9867m_Y  + 0.7685
 =  131.8
 \\
  \w{Var}_X = (0.9867)^2 Var(Y)
 =  408.04
  \end{eqnarray*}
  Note that the observed two first moments of $X$ are given by 131.2 and 439.11.\\
  Similarly  for the pair $(\t{X},\t{Y})$, the parametric predictions are given by

  \begin{eqnarray*}
 \w{m}_{\t{X}}= 0.744 m_{\t{Y}} + 34.787
 = 131.656
 \\
  \w{Var}_{\t{X}} = (0.744)^2 Var( \t{X})
  = 226.933.
  \end{eqnarray*}
The nonparametric approach  yields
 \begin{eqnarray*}
 \w{m}_{\t{X}} = 0.9867m_{\t{Y}}  + 0.7685
 =  129.237,
 \\
  \w{Var}_{\t{X}} = (0.9867)^2 Var( \t{Y} )
 =   399.137
  \end{eqnarray*}
 Recall that the observed two first moments of $\t{X}$ are given by 128.8 and 410.21.\\
  The predictions of the nonparametric model are more close to the observed values, consequently the nonparametric approach seems to be more suitable.

\end{document}